\documentclass[aps,pre,groupedaddress,showpacs]{revtex4}
\usepackage{bm}
\usepackage{bm}
\usepackage{graphicx}
\begin{document}
\title{Oscillatory shear flow in nanochannels via hybrid particle-continuum
  scheme}
\author{R. ~Delgado-Buscalioni}
\email[]{r.delgado-buscalioni@ucl.ac.uk} \affiliation{Centre for 
Computational Science, Dept. Chemistry, University College London, \\
Gower Street, London WC1 OAJ, U.K} 
\author{E. G. Flekk{\o}y} \email[]{flekkoy@fys.uio.no}
\affiliation{Dept. of Physics, Oslo University, \\PB 1048 Blindern, 0316 Oslo, Norway}
\author{P. V. Coveney} \email[]{p.v.Coveney@ucl.ac.uk}
\affiliation{Centre for 
Computational Science, Dept. Chemistry, University College London, \\
Gower Street, London WC1 OAJ, U.K} 
\date{\today}
                           
\begin{abstract}
This paper provides an application of our hybrid continuum-particle
scheme to liquids by considering unsteady shear flows driven by wall
oscillations in nano-slots. The particle region (P) adjacent to the
wall, is described at the atomistic level by molecular
dynamics, while the outer region (C), described by a continuum equation
for the transversal momentum, is solved via a finite volume method. Both
regions overlap in the ``handshaking'' region where a two-way coupling
scheme (C$\rightarrow$P and P$\rightarrow$C) is implemented.  A
protocol for the C$\rightarrow$P coupling was presented in a previous
paper [Delgado-Buscalioni and Coveney, Phys. Rev. E, {\bf 67}, 046704
(2003)];  here we focus on the the P$\rightarrow$C counterpart, that is,
on the exchange of information from the microscopic to
the continuum domain.  We first show how to use the finite volume
formalism to balance the momentum and mass fluxes across the
P$\rightarrow$C surface in such a way that the continuity of velocity
is ensured. Then we analyse the effect of the fluctuations of the (P)
stress tensor on the stability and accuracy of the numerical
scheme. This analysis yields a condition which establishes a
restriction on the resolution of the flow field that can be described by
the hybrid method, in terms of the maximum frequency and wavenumber of
the flow.

\end{abstract}

\maketitle

\section{Introduction}

Many systems, including complex flows near
interfaces,  wetting, drop formation, melting, crystal
growth from a  fluid phase, and moving  interfaces of immiscible fluids
or membranes, are governed by the dynamic interplay between the rapid atomistic processes occurring
within a small localized region and the slow
dynamics occurring within the bulk liquid.
Such problems are too computationally expensive for any standard 
molecular dynamics simulation (MD). A promising alternative is to use hybrid
particle/continuum algorithms which retain the atomistic description
only where it is needed, and reduce the computational cost by 
solving the bulk flow by much faster continuum fluid dynamics methods.

Hybrid particle/continuum algorithms for solids \cite{Abr98} and gases
\cite{Gar00} were the first to be fully developed in the literature.
As  also happens in theoretical descriptions, a hybrid description of
the liquid state is the most challenging one.  The general procedure
is to connect the particle region (P), described by molecular dynamics, 
and the continuum region (C), described by continuum fluid dynamics (CFD), with a 
handshaking region comprised of two buffers: C$\rightarrow$P and
P$\rightarrow$C.  Most of the difficulties encountered when developing
a hybrid scheme for liquids arise at the C$\rightarrow$P buffer, where
the molecular dynamics has to be reconstructed so as to adhere to the
prescriptions coming from the C-flow. Within the P$\rightarrow$C
region the microscopic variables are coarse-grained to supply boundary
conditions for the continuum domain, although within this domain the molecular
dynamics are not directly altered and their motion is 
uniquely governed by the corresponding molecular interactions.

The kind of information to be transferred at the handshaking region
has been the subject of some discussion. The first hybrid schemes to appear
in the literature [see Delgado-Buscalioni and Coveney \cite{BusH1} and references therein]
considered the coupling of momentum in steady shear flows using
schemes based on {\it coupling-through-state}; in particular on
matching the velocity of P and C within the overlapping region. Flekk{\o}y
{\em et al.} \cite{Flek00} introduced another coupling strategy based
on the exchange of the fluxes of conserved quantities (from and to P
and C) and they applied their scheme to transversal momentum and mass
transfer in steady (Poiseuille) flows. This work was subsequently
generalized to include the energy exchange \cite{Flekener}.  More
recently, Delgado-Buscalioni and Coveney \cite{BusH1} introduced
modifications into the flux-based C$\rightarrow$P coupling 
to allow mass, momentum and energy
transfers in time-dependent flows. In that work, the MD
region is surrounded by a continuum description and the C$\rightarrow$P protocol
gives rise to the generalized boundary conditions for the particle system
which provide the correct (hydrodynamic) propagation and relaxation of 
shear, sound and heat waves across the MD region \cite{BusH1}.
The same study compares schemes
based on coupling-through-fluxes and coupling-through-state in the simulation of
a relaxing flow due to decaying pressure waves.  The results showed
that the imposition of the continuity of the thermodynamic variables alone
does not provide the correct decay of the hydrodynamic modes and, 
for example, it yields a negative entropy production \cite{BusH1}. On the contrary, when
using the flux-scheme the correct thermohydrodynamic
behaviour was obtained.

We would like to stress that, although most of the applications that any hybrid method 
must address involve unsteady flows, the previously proposed hybrid schemes for
liquids have barely even considered time-dependent scenarios.
One of the main purposes of this paper is to study the applicability of
the flux-scheme in the simulation of time-dependent shear flows,
ranging from moderate to high characteristic frequencies.
To that end we consider fluid flow in a slot driven by the oscillatory motion
of one of the walls in its own plane (Stokes problem).  This set-up
retains the essential features encountered in unsteady shear flows
while being analytically solvable. 
Oscillatory shear flow is widely
used in rheological studies of complex fluids, such as polymer brushes (see
Wijmans and Smit \cite{Wij02} for a recent review). These systems are good
examples of typical applications of the hybrid scheme, which can
treat the complex fluid region by MD and the momentum transfer from
the bulk by continuum fluid dynamics (CFD). The hybrid 
set-up we are considering here can be also used
in the study of flow induced by complex boundary conditions (such as
rough surfaces) in nanotechnological processes, an active area
of research with important applications in micro electromechanical
systems (MEMS). For instance, Stroock {\em et al.} \cite{Str02} showed that
bas-relief nano-structures can be introduced on the floor of a slot
in order to induce the mixing of solutions in low Reynolds number
flows in microchannels.

Another objective of the present work is to study the effect of
the microscopic fluctuations on the spatial and temporal resolution of the overall
coupling scheme. To that end we quantify and analyze 
the averaging procedure needed in the particle measurements in order
to achieve a given signal-to-noise ratio.
Finally, we consider the velocity discontinuities that arise at the particle-continuum
interface. We show that these can be controlled by using a hybrid velocity
gradient in the imposition of the momentum flux on the continuum system
at the particle-continuum interface. The idea of using hybrid gradients
was first proposed by Flekk{\o}y and Wagner in 
work addressing the hybrid description of the
density field in Fickian diffusion \cite{Flek01}.

The rest of the paper is organized as follows. In Sec \ref{P} we
present the domain decomposition used in our model, give details on the molecular domain
and provide definitions of the spatial and time averages used in this work. 
In Sec. \ref{PC} we present
the hybrid finite volume method used to inject the averaged
microscopic flux into the continuum domain. The protocol for the C$\rightarrow$P coupling
is explained in Sec. \ref{sec.CP}. 
The hybrid scheme produces a finite signal-to-noise ratio 
due to the fluctuation of the particle momentum flux and, in Sec. \ref{FL},
we discuss how to use averaging procedures to maximize it.
In Sec \ref{sec.osc} we apply the hybrid scheme 
to compute oscillatory shear flow in nanochannels. We also
present some hybrid simulations to verify the
conclusions of Sec. \ref{FL} concerning the r\^ole of
fluctuations. Concluding remarks are given in Sec. \ref{Con}.

\section{The molecular region
\label{P}}

\begin{figure}[h]
\includegraphics{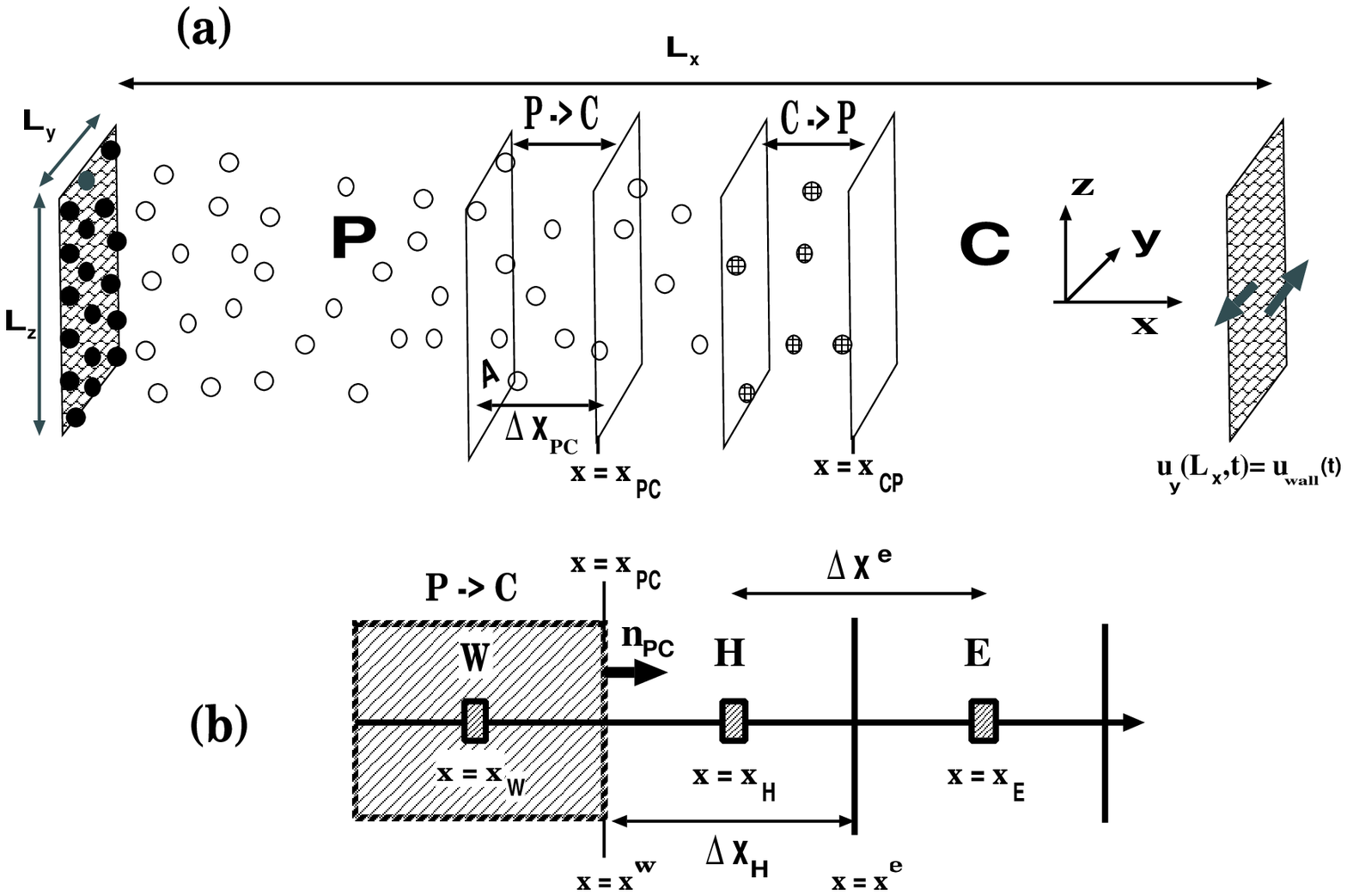}
\caption{ \label{1}
The domain decomposition of the hybrid scheme. (a) displays
the set-up used in the present work (periodic in $y$ and $z$ direction)
indicating the particle domain
(P), the continuum domain (C) and the handshaking domain 
comprised of the C$\rightarrow $P and P$\rightarrow $C cells, whose
surface area is $A$.  The P domain  is contained in $x\leq x_{CP}$ and
includes the wall around $x\sim 0$ formed by two layers of LJ atoms in a
hexagonal lattice (black circles). The C domain is $x_{PC} \leq x \leq L_x$ and includes a
boundary condition at $x=L_x$, which imposes $u_y(L_x,t)=u_{wall}(t)$ (moving wall).
Open circles represent particles whose dynamics
are uniquely determined by their corresponding intermolecular forces. 
Particles entering the C$\rightarrow $P cell (hatched circles)
feel extra forces coming from the continuum domain. (b) shows three
consecutive cells of the finite volume discretization of the C
domain: $H$, $W=H-1$ and $E=H+1$. The index of the cell centres runs from $H=1$ to
$H=M$. For the first cell $H=1$, wherease the west cell $W (=0)$ is outside the C domain
and within the P$\rightarrow $C buffer, as indicated in (b) (see
Sec. \ref{PC}). The centre-to-centre distance is $\Delta x^e=x_E-x_H$
and the face-to-face distance is $\Delta x_H=x^e-x^w$.}
\end{figure}

A typical spatial domain decomposition for our hybrid scheme is 
depicted in Fig. \ref{1}. In the present set-up the overall domain ranges from $x=0$
to $x=L_{x}$ and from $-[L_{\alpha }/2,L_{\alpha }/2]$ in the
other two periodic directions ($\alpha =\left\{ y,z\right\} $). The
particle region (P) spans from $x=0$ to $x=x_{CP}$, the continuum
region (C) from $x=x_{PC}$ to $x=L_x$ and both subdomains overlap within
$x_{PC}\leq x\leq x_{CP}$. 

The particle region (P) contains $N(t)$ 
particles at time $t$, interacting through
interparticle potentials and evolving in time through Newtonian
dynamics. In this work, the particles interact through a truncated
Lennard-Jones (LJ) potential $\psi(r)=\psi_{LJ}(r)-\psi_{LJ}(r_o)$
where $\psi_{LJ}(r)=4\epsilon^{-1}\, \left[(\sigma /r)^{12}-(\sigma
/r)^{6}\right]$. Simulations were performed for the purely repulsive
WCA Lennard-Jones potential \cite{Hes98} using a cut-off radius of $r_{o}=2^{1/6}\sigma$ 
and for the standard LJ fluid using a larger cut-off radius $r_o=3\sigma$.

Each particle has a mass $m$, velocity $\textbf v_{i}$ and energy
$\epsilon _{i}=\frac{1}{2}mv_{i}^{2}+\Sigma _{j}\psi (r_{ij})$
($\mathbf{r}_{ij}=\mathbf{r}_{j}-\mathbf{r}_{i}$). The equations of
motion for the particles are
\begin{eqnarray}
\label{p1}
\dot{\mathbf{r}}_{i}&=&\mathbf{v}_{i} \\ 
\dot{\mathbf{v}}_{i}&=&{\textbf f}_{i}=\sum_{i<j}^{N(t)}r_{ij}^{-1}d\psi (r_{ij})/dr_{ij}\mathbf{r}_{ij},
\label{p2}
\end{eqnarray}
where $\mathbf{f}_i$ is the the force acting on the $i^{th}$ particle.
Equations (\ref{p1}) and (\ref{p2}) were solved via standard molecular dynamics (MD) using the
velocity-Verlet algorithm \cite{allen} with a time step $\Delta t_{P}\simeq
10^{-3}\tau $, where $\tau =(m\sigma ^{2}/\epsilon )^{1/2}$ is the
characteristic time of the LJ potential. Throughout the rest of the
paper, all quantities will be expressed in reduced units of the LJ
potential: $\tau (=0.45\times 10^{-13}$s), $\sigma (=3.305\times
10^{-12}$cm), $\epsilon $, $m(=6.63\times 10^{-23}$g) and $\epsilon
/k_{B}(=119.18$K) for time, length, energy, mass and temperature 
respectively (the numerical values correspond to argon).

The particle system is bounded in the $x$ direction by a lower wall
located at $x=0$ which contains a layer of $1600$ spheres
strongly tethered  to the sites of the $(1,1,1)$ plane of an fcc lattice
by harmonic springs of stiffness $\kappa=1320 \epsilon
\sigma^{-2}$. The wall atoms do not interact with each other and the
wall-fluid interaction is LJ with an increased cutoff and potential well,
$r_{o,w}=1.311\sigma$ and $\epsilon_{w}=1.303$. These values ensure
vanishingly small slip velocities for the shear rates considered here.

In order to maintain  the temperature of the system constant we used
either a Nos\'e-Hoover thermostat \cite{Frenkel.book} for the LJ fluid,
or a Langevin thermostat \cite{robbins,Frenkel.book} for the WCA-LJ
fluid. The local kinetic temperature,
needed for the Nos\'e-Hoover thermostat, was
calculated from the peculiar velocities
within slices of width $1.0 \sigma$ along the $x$
direction.  Also, to ensure that the Langevin 
thermostat does not bias the shear profile, the Gaussian white
noise and damping terms are only added to the equations of motion for
the velocity components normal to the mean flow, the $x$ and $z$
directions  \cite{robbins}. Both thermostats proved
to be equally efficient.

The coupling scheme is a two-fold communication between C and P, as
can be understood from Fig. \ref{1}. Within the C$\rightarrow $P domain, the
fluxes evaluated from the continuum solution are imposed on the
particle dynamics. This part of the coupling scheme has been discussed
by Delgado-Buscalioni and Coveney \cite{BusH1} for the general
scenario of mass, momentum and energy transfer in unsteady relaxing
flows. On the other hand, the flux due to the particle
dynamics across the P$\rightarrow$C interface is to be injected into the
continuum system as flux boundary conditions. 
In order to coherently communicate information between both
descriptions of matter, the fluxes arising from the particle system 
must be averaged in time and space prior to transferring to the
coarse-grained level. Moreover these
averages need to be local on the coarse-grained coordinates
and on the time scale of the C domain. To that end,
around each  coarse-grained spatial coordinate, $\textbf R$, 
we define a cell of volume $V({\bf R})$ and, for any
particle variable, say $\phi({\bm r}_i,t)$, we define the local mean
$\Phi({\bm R},t)$ at the coarse grained location $\textbf R$ as
\begin{equation}
\label{sums}
\Phi({\bm R},t)=\frac{ \int_{V({\bf R})} \sum_{i=1}^N \phi({\bf r}_i,t) \delta({\bf r}-{\bf
  r}_i) d^3{\bf r}}{\int_{V({\bf R})} \sum_{i=1}^N \delta({\bf r}-{\bf  r}_i)
  d^3{\bf r}},
\end{equation}
where the integration is performed over the cell volume $V({\bf R})$ and
the summation runs over the particle index. The corresponding 
spatio-temporal average is also local in the ``coarse-grained'' time
$t_C$ with a time averaging window of length $\Delta t_{av}$,
\begin{equation}
\label{sum}
\left<\Phi \right>(\mathbf{R},t_{C})=\frac{1}{\Delta t_{av}} \int _{t_{C}-\Delta t_{av}}^{t_{C}} \Phi ({\bf R},t) \delta (t-t_{k})dt.
\end{equation}
For the temporal average a number of samples $n_{s}$ of the microscopic
quantity are taken along the time interval $\Delta t_{av}$ at times
given by $t_{k}=t_{C}-\Delta t_{av}+k\delta t_{s}$ ($k=\left\{ 1,...,n_{s}\right\} $).
To ensure statistically independent measurements the time interval between
samples $\delta t_{s}$ cannot be smaller than the decorrelation time
of $\Phi$, which can be calculated from $\int_0^{\infty}
\langle\Phi^{\prime}(t)) \Phi^{\prime}(0) \rangle dt/\langle
(\Phi^{\prime})^2\rangle$, where $\Phi^{\prime}\equiv \Phi-\langle
\Phi \rangle$ (see Sec. \ref{FL}).

As shown in Sec. \ref{PC}, the average defined by Eqs. (\ref{sums}) and
(\ref{sum}) is used within the
P$\rightarrow$C cell to evaluate the coarse-grained momentum flux from
the particle region to the continuum domain. In
this case the spatial average is computed within the volume 
$V_{PC}\equiv V({\bf R}_{PC})$, where ${\bf R}_{PC}$ is the 
location of the P$\rightarrow$C cell.
This average shall be denoted  $\langle \Phi\rangle_{PC}(t)\equiv
\langle \Phi\rangle({\bf R}_{PC},t)$. We 
shall focus on the P$\rightarrow $C coupling of the shear stress.
At any given time instant, the time-averaged mean momentum flux tensor evaluated within 
the P$\rightarrow $C cell is given by
\begin{eqnarray}
\label{jp}
\langle \bm{j}_{p}\rangle & = & \frac{1}{V}_{PC}\left<\Sigma_{i=1}^{N_{PC}}m\bm{v}_{i}\bm{v}_{i}-\frac{1}{2}\Sigma_{i,j}^{N_{PC}}\bm{r}_{ij}\bm{F}_{ij}\right>\cdot \bm{n}_{PC},
\end{eqnarray}
where  $N_{PC}$ is the number of particles inside the P$\rightarrow $C
cell, $\rho_{PC}=N_{PC}/V_{PC}$, and $\mathbf{n}_{PC}$ is the surface vector
of the cell pointing toward the neighbouring C-region (see Fig. \ref{1}b).

\section{Continuum description and P$\rightarrow $C coupling
\label{PC}}

Within the C region the relevant variables are the macroscopic local
densities associated with the conserved quantities. If $\Phi $ is
any conserved quantity (per unit mass), its conservation law is given
by \begin{equation}
\frac{\partial \rho \Phi }{\partial t}=-\nabla \cdot {\bf J_{\Phi }},\label{Cflow}\end{equation}
In this work we shall consider isothermal fluids and restrict ourselves
to the momentum density $\rho \bf {u(\bf {r,t)}}$. In what follows,
the continuum velocity will be denoted by 
${\bf u}$ while $\langle {\bf v}\rangle$ stands for the mean particle velocity.
The flux of momentum is given by $\mathbf{J}_{\bf u}=\rho \mathbf{uu}+\bm \Pi $, where the
pressure tensor $\bm \Pi =P\, \mathbf{1}+\bm \tau $ includes the
local hydrostatic pressure $P$ and the viscous stress
tensor. 
For an LJ fluid, the latter satisfies a Newtonian constitutive
relation \cite{Cic84,Hey88}, 
$\bm \tau =-\eta \left( \nabla \bm{u}+ \nabla \bm{u}^{T}\right)
+\left(2\eta /3-\xi \right)\nabla \cdot \bm{u}$. where $\eta$ is the
shear viscosity and $\xi$ the bulk viscosity.
In principle, the set of equations given by Eq.(\ref{Cflow}) can be
solved by any standard continuum fluid dynamics (CFD)
method. Nevertheless, as the
hybrid scheme proposed here is based on the balance of fluxes
it fits naturally with the finite volume method \cite{Patankar}, which
exactly balances the fluxes across the computational cells. 

In order to illustrate the P$\rightarrow$C coupling protocol we shall
consider the flow of an incompressible and
isothermal fluid with mean density $\rho_c$ and mean temperature
$T_c$. The fluid fills the space between two parallel walls at $x=0$
and $x=L_x$. The wall at $x=L_x$ is  moving with arbitrary velocity
$u_{wall}(t)$ along the $y$ direction. The flow
is uniquely driven by the motion of this wall,
meaning that the mean pressure is constant 
throughout the domain, $P=P_c(\rho_c,T_c)$, 
and that there are no transfers of mean energy or
mass along the $x$ direction (perpendicular to the P$\rightarrow$C
interface).  As discussed by Vogel {\em et al.}  \cite{Vog03},
if the system is not confined
in the flow direction (as happens to be the case in our system)
this flow is laminar and the velocity of the continuum flow is given 
by ${\bf u}=u_y(x,t) {\bf j}$. The stress tensor is 
${\bf J_u}=P_c\,{\bf 1}- \eta \partial u_y/\partial x  {\bf j}$.
In the case of the LJ fluid, the value of the dynamic viscosity $\eta$
was obtained from Heyes \cite{Hey88} and for the WCA-LJ fluid 
$\eta$ was measured from MD simulations 
using a standard non-equilibrium procedure \cite{Evans}, i.e.,
from the ratio of the shear stress and shear rate in Couette flow. The equation of motion 
for the $y$-momentum is given by the unsteady diffusion equation
\begin{equation}
\label{eqvy}
\frac{\partial u_y}{\partial t} = \nu \frac{\partial^2 u_y}{\partial x^2},
\end{equation}
where $\nu=\eta/\rho_c$ is the kinematic viscosity. At the fixed wall,
the boundary condition is $u_y(0,t)=0$ and at the moving wall
$u_y(L_x,t)=u_{wall}(t)$. 

For the finite volume discretization of Eq. (\ref{eqvy}) in the
geometry of Fig. \ref{1} we divide the extent of the C region along
the $x$-coordinate ($x_{PC}\leq x\leq L_x$) into $M$ cells. The faces of
the cells can be placed at arbitrary positions $x_k^{f}$ with
$k=\left\{0,...,M\right\}$, and the centres of the cells at
$x_k=(x^{f}_k+x^{f}_{k-1})/2$ with $k\in\left\{1,...,M\right\}$.  To
derive a closed set of equations for the velocities $u_y(x_k)$ at the
$k\in\left\{1,..,M\right\}$ cell centres, one integrates
Eq. (\ref{eqvy}) over the volume of each cell. For a given cell, say
$k=H$, the integration is done over $x^w\leq x\leq x^e$, where the
cell faces of the $H$ cell ($x^w$ and $x^e$) are illustrated in
Fig. \ref{1}b.  The resulting first order spatial derivatives at the
cell faces $x^e$ and $x^w$ are discretized as, e.g., $\partial
u_y(x^e)/\partial x =(u_y(x_{E})-u_y(x_{H}) )/(x_E-x_H)$; where the
``east'' cell $E=H+1$ is placed at $x_E=x_{H+1}$ and the ``west'' cell
$W$ at $x_W=x_{H-1}$.  We use an explicit time integration scheme by
approximating the time derivative at the centre of the cell by
$\partial u_y/\partial t \simeq (u_y(t +\Delta t_C)-u_y(t))/\Delta
t_C$, where $\Delta t_c$ is the time step.  We refer to the excellent
book of Patankar \cite{Patankar} for a comprehensive presentation of
the finite volume discretization of Eq. (\ref{eqvy}) and Navier-Stokes
equations including a discussion on the different time integration
schemes. The resulting discretized version of equation (\ref{eqvy})
reads,
\begin{eqnarray}
\label{exp}
u_y(x_{H}, t+ \Delta t_C) &=& (1-r_e-r_w)\,u_y(x_{H},t)+r_e\,u_y(x_{E}, t)+r_w\,u_y(x_{W},t), \\
\nonumber &&\mathrm{with}\;E=H+1,\,W=H-1\; \mathrm{and}\;H=\left\{1,..,M\right\}.
\end{eqnarray}
We have introduced $r_e$ and $r_w$, defined as $r_f\equiv \nu \Delta
t/(\Delta x^f \Delta x_H)$ where $f=\left\{w,e \right\}$, $\Delta
x^f$ is the the centre-to-centre distance and $\Delta x_H$ the face-to-face distance,
as shown in Fig. 1b.  Most of the simulations reported in this work were
performed with a regular grid \footnote{Note that for a regular grid
the finite volume discretization of Eq. (\ref{eqvy}) coincides with
the explicit finite difference version.} ($r\equiv r_w=r_e$), using
$\Delta x \sim 0.5$.   The time step $\Delta t_C$ was chosen to fulfil the
Courant condition $r\leq 1/2$, which ensures the stability of the
explicit scheme in Eq. (\ref{exp}) \cite{Patankar}.
Note that Eq. (\ref{exp}) is not closed for the velocities at the
$H=\left\{1,..,M\right\}$ cell centres. For instance, the evaluation of the
velocity at the cell $H=M$ requires the
velocity at the outer cell $E=M+1$, which is 
outside the computational domain of the C region.
Similarly, for the velocity at $H=1$ one requires the velocity at the outer cell
$W=0$. The velocities at these outer cells (also called {\em ghost} cells) have
to be evaluated from the boundary conditions. For instance,
the velocity of the {\em ghost} cell $E=M+1$ at $x_{M+1}=L_x+\Delta
x_E/2$ is calculated by linear extrapolation from the velocity at the 
$H=M$ cell and the velocity at the wall:
$u_y(x_E,t)= u_y(x_{H},t) + 2 \Delta x^e\,
(u_{wall}(t)-u_y(x_{H},t))/\Delta x_H$, where $H=M$.

In the same way, to close Eq. (\ref{exp}) for $H=1$ we need to determine the velocity 
at the {\em ghost} cell $W=0$ at $x_0=x_{PC}-\Delta x_W/2$. 
To that end we use the hybrid formulation and
set the $y$-momentum flux across P$\rightarrow$C equal to
the time-averaged momentum flux due to the particle system at the
P$\rightarrow$C cell: $\langle j_{p,xy}\rangle_{PC} =-\eta (\partial u_y/\partial x)_{x_{PC}}$. 
The discretized version of the right hand side (RHS) of this momentum flux balance equation 
is expressed as $-\eta ({\bar u_y}(x_{H})-u_y(x_{W}))/\Delta x^w$, with
$H=1$ and $W=0$. Note that both the particle and the continuum domains
overlap at the first cell $H=1$, so the velocity
${\bar u_y}(x_{H})$ could be chosen to be either the time averaged 
mean particle velocity at the $H=1$ cell, $\langle v_y\rangle(x_H)$
or the continuum velocity at the same location, $u_y(x_H)$. The
implication of such a choice will become clear shortly. We have
enabled a set of possible choices based on a linear combination of both velocities,
\begin{equation}
\label{upb}
{\bar u_y}(x_{H})=(1-\alpha )u_y(x_{H})+\alpha \langle v_y\rangle(x_{H}),\;\;\mathrm{for}\; H=1\;  \mathrm{and}\; \alpha\in[0,1].
\end{equation}
The flux balance equation $\langle j_{p,xy}\rangle_{PC}=-\eta({\bar
u_y}(x_{H})-u_y(x_{W}))/\Delta x^w$ provides the velocity at the
ghost cell $u_y(x_W)$, required to close Eq. (\ref{exp}) for $H=1$,
\begin{equation}
\label{uw}
u_y(x_W)={\bar u_y}(x_H)+\langle j_{p,xy}\rangle_{PC} \Delta x^w/\eta
 \;\;\mathrm{for}\;W=0 \; \mathrm{and}\; H=1.
\end{equation}
Insertion of Eq. (\ref{uw}) and Eq. (\ref{upb}) into Eq. (\ref{exp}) provides the
time advancement algorithm for the velocity at the cell interfacing the
particle system:
\begin{eqnarray}
\label{up}
u_y(x_{H},t+\Delta t_C)&=&(1-r_e)\,u_y(x_{H},t) +r_e\,u_y(x_E,t)
+\frac{\langle j_{p,xy}\rangle_{PC}(t) \Delta t}{\rho \Delta x^w}+\\
\nonumber &+&
\alpha r_w\,\left(\langle v_y\rangle(x_{H},t)-u_y(x_{H},t)\right),\;\;\;
\mathrm{for} \; H=1, E=2.
\end{eqnarray}

\begin{figure}[h]
\includegraphics[width=12cm,totalheight=10cm]{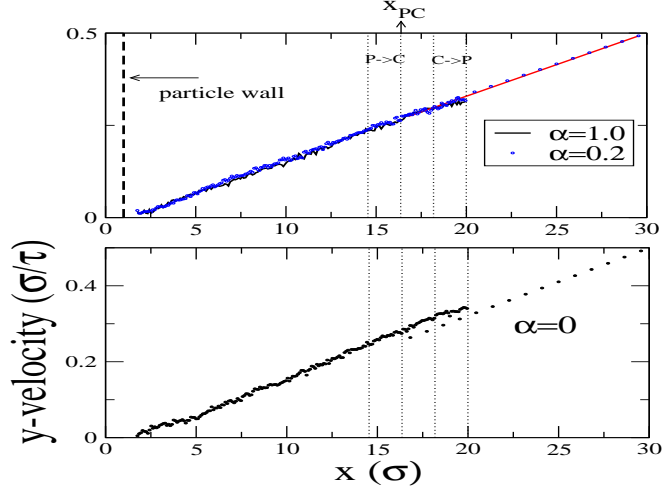}
\caption{The $y$-velocity versus the $x$ coordinate for tests done in a Couette flow  with wall velocity
$u_{\max}=0.5$.  The  particle region comprises
$x \leq 20$ and the continuum domain is $x\in [16.36,30]$.  The mean
density  is $\rho  =0.8$ and  the temperature  is $T=1.0$. 
Vertical lines indicate the average $x$ position of particle wall $(x=1)$,
the P$\rightarrow$C interface $x_{PC}=16.31$ and the extent of 
the P$\rightarrow$C and C$\rightarrow$P cells, $\Delta x_{PC}=1.81\sigma$.
The upper figure shows the $y$-velocities averaged over $\sim 800 \tau$
obtained for $\alpha=1$ and $\alpha=0.2$. The lower figure shows the
$y$-velocities obtained using $\alpha=0$, averaged over $\sim 700 \tau$.}
\label{alfavy}
\end{figure}

\begin{figure}[h]
\includegraphics[width=12cm,totalheight=10cm]{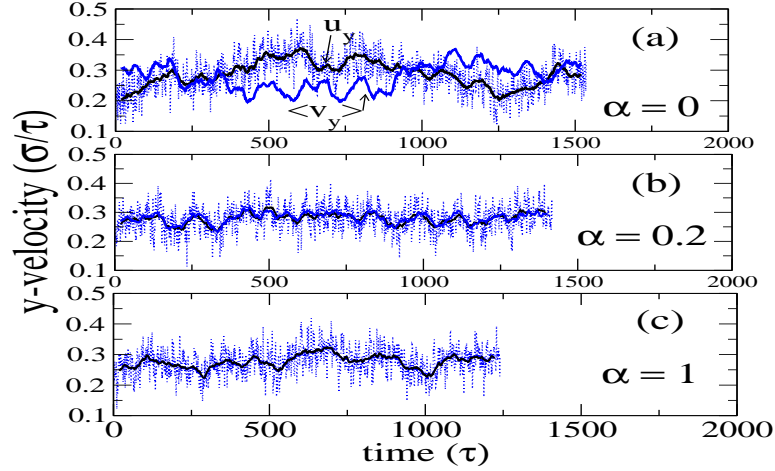}
\caption{The  effect of the  $\alpha$  parameter in  Eq. (\ref{up}) on  the
continuity of the  $y$-velocity at $x_1=x_{PC}+\Delta x/2$
($x_{PC}=16.31$ and $\Delta x=0.9$) for the same Couette flow tests shown in Fig. \ref{alfavy}.
From top to bottom the figures correspond to $\alpha=0$, $\alpha=0.2$ and
$\alpha=1$.
The continuum  velocity at  $x=x_1$ is labelled  by $u_y$  and the mean
particle velocity within the region $|x-x_{1}|\leq 0.9$ is denoted as
$\langle v_y \rangle$.  Dotted  lines  correspond to  velocities
averaged   over  a short time interval $\Delta t_{av}=1\tau$, and  solid lines are the
same velocities averaged over $20\tau$.
The velocities $u_y$ and $\langle v_y \rangle$ are indicated in (a), while 
in (b) and (c) their values are almost identical.  
Note that the size of the velocity discontinuity observed for $\alpha=0$ is
similar to the amplitude of the velocity fluctuations (dotted lines).
\label{alfav}}
\end{figure}

The effect of the choice of $\alpha$ in Eq. (\ref{upb}) becomes clear
by inspection of Eq. (\ref{up}).  For any $\alpha \ne 0$ the last term
on the RHS of Eq. (\ref{up}) acts as a relaxation term that ensures 
velocity continuity by gently driving the continuum velocity at the
boundary cell $x=x_1$ to the corresponding particle average $\langle
v_y\rangle(x_1)$.  This term vanishes or becomes negligible
once velocity continuity is established, $u_y(x_{1})\simeq \langle
v_y\rangle(x_{1})$.  We emphasise that this sort of {\em velocity
coupling} does not directly act on the particle dynamics, but only within
the algorithm for the continuum dynamics. This fact is significant because
it avoids any artificial alteration (such as a Maxwell d{\ae}mon) to the
microscopic dynamics of the particles near the P$\rightarrow$C
interface.  The idea of using the scheme in Eq. (\ref{up}) arose from the outcome
of calculations performed at low shear rates for which $du_y/dx <10^{-2}$. 
At low shear rates the velocity
fluctuations become significant enough to produce deviations of the 
mean particle and continuum velocities at the overlapping region.
If the velocity coupling term is absent in Eq. (\ref{up}) ($\alpha=0$) the
deviations of the P and C velocities produced by subsequent particle
fluctuations can be maintained in time, although the slopes of the P
and C velocities do converge to the same value 
(see Figs. \ref{alfavy}, \ref{alfav} and discussion
below).  The reason for this fact is clear: the imposition of a flux does
not prescribe the value of the velocity, but only its  gradient.
As shown below, the algorithm in Eq. (\ref{up}) solves the problem of velocity
continuity even when a small velocity coupling is used $\alpha$
 $(\sim 0.1)$.

To demonstrate the effect of the parameter $\alpha$ 
in Eq. (\ref{up}) on velocity continuity
we performed some hybrid simulations of Couette flow 
using a Lennard-Jones fluid with density $\rho=0.8$
and temperature and $T=1.0$. Similar tests were also done in the
simulations of oscillatory shear flow presented in Sec \ref{sec.osc}.
These tests were performed with $\alpha=\left\{0,0.1,0.2,1\right\}$. 
Figure \ref{alfavy} shows the $y$-velocity profile
of a Couette flow with $du_y/dx=0.01724$, obtained by averaging within slices
of width $0.2\sigma$ and over the whole simulation time.
Results obtained for $\alpha>0$ 
do not present any velocity discontinuity and are in perfect agreement with the analytical
profile (the local shear rate at the $x_{CP}$
interface deviates by less than $2\%$ from $u_{wall}/L_x$).  On the
contrary, when using $\alpha=0$ in Eq. (\ref{up}), the averaged
velocity profile presents a significant discontinuity at the
overlapping region which affects the overall shear rate of
the flow.  As shown in Fig. \ref{alfavy} the discontinuity tends to be
reduced as the average time is increased, but it is still significant
after an average time over $1500 \tau$.

The effect of the parameter $\alpha$ in the continuity of velocity is
more clearly shown in Fig. \ref{alfav} where the continuum and mean
particle velocities at the boundary cell $H=1$ are plotted {\em versus}
time. Using $\alpha=1$ the relative differences
between the C and P velocities are less than $2\%$, while for a value
of $\alpha$ as small as $0.2$, these relative differences are still
less than $5\%$. By contrast, when using $\alpha=0$, this relative
difference increases to $\sim 80\%$.  As seen
in Fig. \ref{alfav}, for $\alpha=0$ the C velocity and averaged P
velocity at the cell $H=1$ ``oscillate'' around the correct value
$\sim 0.28$, and the period of these oscillations can be as large as
$\sim 600 \tau$. For smaller shear rates the situation worsens because this
period may become much larger. In summary, using $\alpha=0$
the numerical scheme of Eq. (\ref{up}) does not guarantee
velocity continuity. Inspection of Fig. 3a clearly indicates that 
using $\alpha>0$ enables us to deal with
time-dependent flows, as shown in Sec. \ref{sec.osc}.

The central idea of our 
hybrid scheme is the balance of fluxes and such a balance must to be
respected by any choice of $\alpha$. 
To investigate this we compared the time averaged flux injected from the particle
system with the time averaged flux measured from the continuum field at the interfacing
cell, $H=1$. If the flux exchange is correctly balanced, both quantities
should coincide. The outcome of these tests
is shown in Table 1. For any $\alpha\in[0,1]$, both fluxes 
differ by less than $2\%$ indicating that the velocity coupling term 
in Eq. (\ref{up}) does not alter the balance of fluxes.

\begin{table}[h]
\begin{tabular}{c|cc||cc|}
    & Particle & system & Continuum & system  \\
$\alpha$ & $\langle j_{p,xy} \rangle_{PC}$       & $S_{j_p}$ &  
           $\Pi_{xy}(x_{PC})$ &  $S_{\Pi}$ \\
\hline
0   & 0.0289 & 0.09 & 0.0294 & 0.01 \\
0.2 & 0.0287 & 0.08 & 0.0294 & 0.01 \\
1   & 0.0298 & 0.08 & 0.0297 & 0.01 \\
\end{tabular}
\caption{The fluxes across the P$\rightarrow$C interface for different
values of the $\alpha$ parameter. Comparison is made between the time averaged
particle flux $\langle j_{p,xy} \rangle$ 
and the time averaged momentum flux in the
continuum field, $\Pi_{xy}(x_{PC})\equiv-\eta (du_y/dx)_{PC}$. 
Time averages were performed over $\sim 1000\tau$.
The results were obtained from a Couette flow with
imposed shear rate $u_{wall}/L_x=0.0172$ (see Fig. \ref{alfavy}).
The momentum flux measured from the C flow is calculated using $\eta=1.75\pm
0.05$ and a discretized version of the velocity gradient near the
P$\rightarrow$C interface, $du_y/dx=(u(x_2)-u(x_1))/\Delta x$
(where $x_1=x_{PC}+\Delta x/2$, $x_i=x_1+i\Delta x$ and 
$\Delta x=0.9$ is the width of the spatial mesh). For all values of $\alpha$ 
the time averaged P and C-fluxes coincide well within their fluctuation
margins and agree with the externally imposed stress, $\eta
u_{wall}/L_x =0.030\pm 0.001$. The amplitude of the fluctuations 
of both fluxes (their standard
deviation) has been also indicated by $S_j$ and $S_{\Pi}$. All
quantities are expressed in LJ reduced units $m\sigma^{-1}\tau^{-2}$. }
\end{table}

We note that the velocity coupling term in Eq. (\ref{up}) 
introduces an extra momentum flux into the C system whose magnitude
is $\alpha \eta \left(\langle v_y\rangle(x_{H},t)-u_y(x_{H},
t)\right)/\Delta x$.  This flux fluctuates on the microscopic time
scale and its average is close to zero. This fact explains why the
relaxing term does not affect the flux balance on the time scale of
the coarse-grained dynamics, as can be seen in Table 1.  
We measured the relative size of the flux contributions in
Eq. (\ref{up}) and concluded that 
the size of the relaxing term remains much smaller than the particle flux 
for shear rates larger than $\gamma >10^{-3}$.
Below a certain shear rate, $\gamma\sim 10^{-3}$, the fluctuations of the 
relaxing term are of the same order as the particle flux signal. 
Nevertheless, for such small shear rates the signal-to-noise ratio of the particle flux becomes
larger than one (see Sec. \ref{FL}) and we are obliged to perform
temporal averages over long times to recover the correct 
velocity profile and flux balance. We performed simulations of a Couette flow at
$\gamma=10^{-3}$ using $\alpha=0.5$. The values of the fluxes, averaged
over a time of $10^4\tau$, were $\langle j_{p,xy} \rangle_{PC}=(1.9\pm 2.1)\times 10^{-3}$
while $-\eta (du_y/dx)_{PC}=(1.8\pm 0.3)\times 10^{-3}$ (here the error bars
denote half the standard deviation). These values compare well with
the externally imposed stress, $\eta u_{wall}/L_x=(1.75\pm
0.05)\times 10^{-3}$. In passing, we note that 
owing to the effect of the strong fluctuations, when simulating flows with 
$\gamma\sim 10^{-3}$ using $\alpha=0$, the averaged P and C velocities in the
handshaking region never matched; hence, although the fluxes (and therefore the
velocity gradients) were perfectly balanced, they did not agree with
the desired, externally imposed, value.

One might ask what value of $\alpha \ne 0$ is best
suited for a certain simulation. We give here indications as to how to answer this question.
Note that the velocity coupling takes a characteristic
relaxation time $O[\Delta x^{2}/(\nu\, \alpha)]$ to drive the
continuum velocity towards the average particle velocity.  As stated
above this time can be very short, for instance in simulations with
$\alpha=1$, $\Delta x \sim 1$ and $\nu \sim 2.0$, one finds a relaxation time  $\sim 1
\tau$ which lies within the characteristic time of the microscopic
fluctuating currents \cite{BusH1}. In flows where the
amplitude of particle fluctuations is expected to be large (e.g. at
very low shear rates or near the critical point) it may be desirable
to increase this relaxation time by using a smaller value of
$\alpha\sim 0.1$. In this way one can prevent large jumps in the
continuum velocity which may lead to numerical instabilities of the
CFD algorithm for the continuum domain.

\section{The C$\rightarrow$P coupling: control of mass fluctuations
\label{sec.CP}}
For a detailed description of this part of the hybrid scheme we refer to
Delgado-Buscalioni and Coveney \cite{BusH1}, where
the C$\rightarrow$P protocol is applied to the general
scenario of flows with time-dependent mass, transversal and momentum
and energy transfers. For completeness we now briefly explain how the momentum flux
from the C flow is injected into the particle system. We also present
a way to control the fluctuations of mass across the
C$\rightarrow$P interface so as to ensure the balance of mass flux.

\subsection{Imposition of the shear stress on the particle system}
\begin{figure}[h]
\includegraphics[width=8cm,totalheight=15cm,angle=-90]{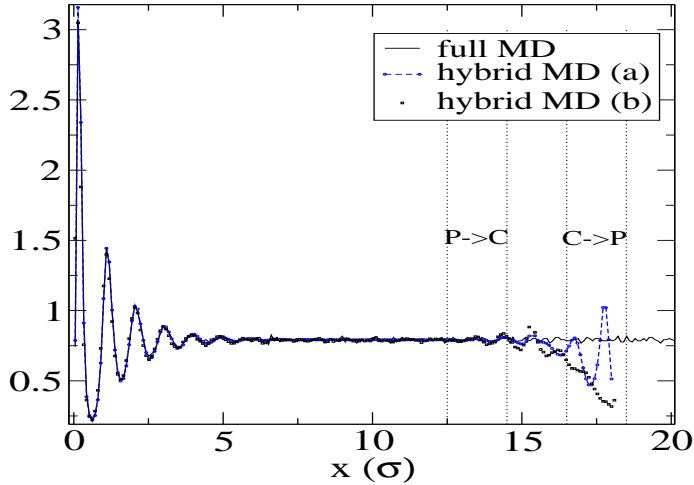}
\caption{The particle density versus the direction normal to the atomistic 
wall at $x\leq 0$ for a WCA-LJ fluid of density $\rho_c=0.8$ and temperature $T=1.0$.
The density profile obtained with the hybrid scheme is compared with the
outcome of a full MD simulation with a second atomistic wall at
$x=50\sigma$. (a) corresponds to $\rho_O=0.7$ in Eq. (\ref{mfl}) and (b)
to $\rho_O=0.5$. The width of the C$\rightarrow$P cell is $\Delta x_{CP}\simeq 2.3\sigma$.
The locations of the P$\rightarrow$C and C$\rightarrow$P cells are also shown
(see Sec. \ref{CP} for further details). \label{dens} }
\end{figure}

This part of the coupling scheme occurs
within the C$\rightarrow$P cell shown in Fig. \ref{1}a.
The stress induced by the C-flow in the P domain is given by the local
momentum flux at the C$\rightarrow$P interface at $x=x_{CP}$.
The flow considered here only carries transversal momentum 
(i.e. along the $y$-direction), so the momentum flux is given by
\begin{equation}
{\bf \Pi}_{CP}\cdot{\bm n}_{CP} =-P_c {\bf i} +\eta (\partial u_y/\partial x)_{x_{CP}} {\bf j},
\end{equation}
where ${\bm n}_{CP}=-{\bf i}$ is the surface vector of the
C$\rightarrow$P interface.  In order to introduce this stress into the
particle dynamics we add an overall external force ${\bf F_{ext}}=A
{\bf \Pi_{CP}}\cdot{\bf n}_{CP}$ to those particles within the
C$\rightarrow$P cell.  For any instant of time, $t$, this force is
equally distributed among the $N_{CP}(t)$ particles inside the
C$\rightarrow$P cell, so the external force per particle is ${\bf
F_{ext}}/N_{PC}=A\,{\bf \Pi_{CP}}\cdot{\bf n}_{CP}/N_{CP}$.  Note that
this external force has a component normal to the C$\rightarrow$P
interface, which provides the hydrostatic pressure, and a tangential
component providing the shear stress.  The particles are free to enter
or leave the C$\rightarrow$P region, so the number of particles within
this region $N_{CP}(t)$ and the value of the overall external force
fluctuate in time. The mean value of $N_{CP}$ is $A\Delta x_{CP}
\rho_{CP}$ so the average ``pressure force'' per particle is
$P_c/(\Delta x_{CP} A^2 \rho_{CP})$, where $\Delta x_{CP}\simeq
2\sigma$ is the extent of the C$\rightarrow$P cell along the $x$
direction, $A= L_y \times L_z$, the local density is
$\rho_{CP}=N_{CP}/V_{CP}$, and the pressure $P_c(\rho_c,T_c)$ is given
from the equation of state which in the case of the WCA-LJ fluid was
provided by Hess {\em et al.} \cite{Hes98} and for the standard LJ
fluid by Johnson {\em et al} \cite{eosLJ}. This normal force prevents
the escape of particles and maintains the correct value of the density
across the inner part of the MD domain as seen in
Fig. \ref{dens}.  The shear force is distributed over the
particles in the same way as described above for the pressure force.
In this case, the flux of $y$-momentum to be injected in the particle
system is $\eta (\partial u_y/\partial x)_{x_{CP}}$.

\subsection{Control of mass fluctuations across the C$\rightarrow$P
  interface \label{CP}}
For the flows we are treating here, the mean fluxes of mass and energy across the C and P
interfaces are zero but fluctuations in the particle system
produce perturbative mass currents along the $x$ direction which need to be
controlled. For this purpose, we coupled the end region of the P domain with a
control equation that governs the mass flux across the particle system frontier. 
This control equation is derived using the finite volume rationale: 
we integrate the mass conservation equation $\partial \rho/\partial
t=-\nabla (\rho {\bf u})$ in a control 
volume around the C$\rightarrow$P interface (between $x_{CP}=x_{CP}-\Delta x_{CP}/2$ 
and $x_O\equiv x_{CP}+\Delta x_{CP}/2$) to
obtain the particle flux towards the C domain.
Zero mass flux is ensured by equating this rate to the
rate of insertion in the particle system, $\dot N_{PC}$. This yields
\begin{equation}
\label{s1}
\dot{N}_{CP}= A \left(\langle \rho v_x\rangle_{CP}- \rho_c u_x(x_0) \right) =
 A \langle \rho v_x\rangle_{CP},
\end{equation}
where $\dot{N}_{CP}$ is the time derivative of the number of particles
within the C$\rightarrow$P cell. The mean particle mass flux $\langle \rho v_x\rangle_{CP}$ is evaluated within
the C$\rightarrow$P cell and averaged over $\Delta t_{av}$.  On the
right hand side of Eq. (\ref{s1}) we have used the fact that the normal
velocity in the C domain is zero, $u_{x}(x_0)=0$.  Equation (\ref{s1}) proved to be
successful in ensuring mean zero mass flux and in eliminating any incoming
longitudinal mass currents towards the C$\rightarrow$P interface.

We also used a different strategy to fix
the rate of particle insertion/removal, which provides further control on the particle density
near the C$\rightarrow$P interface. In this second approach 
the particle insertion rate is chosen to make the particle number in the
C$\rightarrow$P cell $\rho_{CP}$ relax to a pre-set value $\rho_O$,
\begin{equation}
\label{mfl}
\dot{N}_{CP}=\frac{V_{PC}}{\tau_{r}} \left(\langle \rho \rangle_{CP} -\rho_{O} \right),
\end{equation}
where $V_{PC}$ is the volume of the C$\rightarrow$P cell 
and $\tau_{r}$ is a relaxation time. As before, $\langle .\rangle$
denotes the time average over $\Delta t_{av}$. The
amplitude of the density fluctuations within the C$\rightarrow$P cell is
controlled by the coefficient $\tau_{r}$ and can be chosen according
to the requirements of the problem. For instance, 
in hybrid simulations of
tethered polymers under Couette flow \cite{SandraBus} one may need to
smooth out the perturbative density waves originated by the tumbling
motion of the polymer at the inner part
of the MD domain. Otherwise these waves can bounce back at the
C$\rightarrow$P interface and affect the dynamics of the polymer. To
that end, the relaxation time $\tau_r$ need to be smaller than the
time needed for a sound wave to cross the width of the C$\rightarrow$P
buffer, $\Delta x_{CP}/c_s$, where $c_s$ is the sound velocity 
(see e.g. \cite{BusH1}).
The procedure implemented according to Eq. (\ref{mfl}) is an artifact that ensures 
fluctuations carrying mass and longitudinal currents are filtered
out from the simulation
box. In problems where the pressure waves are an essential part of the
study, the mass flux will be governed by the longitudinal
momentum and energy equations, which will need to be added into the hybrid formalism.
According to Eq. (\ref{mfl}), particles are extracted if $\dot
N_{PC}<0$ and, as explained in Ref. \cite{BusH1}, the first particles
to be extracted are those closest to the C$\rightarrow$P interface.  If
$\dot N_{PC}>0$, new particles are inserted with a velocity extracted
from a Maxwellian distribution with mean velocity 
and temperature given by the continuum values ${\bf u}(x_{CP})$ and $T=T_c$.  
The insertion of particles in
dense fluids is a far from trivial task but it has been solved by
the {\sc usher} algorithm proposed
by Delgado-Buscalioni and Coveney \cite{usher}.  

The value of $\rho_O$ in
Eq.(\ref{mfl}) was set to a slightly smaller value than 
the mean density. The reason for this choice was first to alleviate
the computational cost of insertion 
\footnote{As shown in
Ref. \cite{usher},  to insert an LJ atom at
a location where the potential energy equals the mean specific
potential energy of the system
the {\sc usher} algorithm needs around 30 iterations if the mean
density is $\rho_c \simeq 0.8$, and 15 iterations if $\rho_c \simeq 0.7$
(each iteration corresponding to the evaluation of a single-particle
force).} and second to reduce the amplitude
of ripples of the density profile near the C$\rightarrow$P buffer.
For a fluid with $\rho_c=0.8$, we compared the density profile 
arising from a full MD simulation
with that resulting using $\rho_O=0.7$ and
$\rho_O=0.5$ in Eq. (\ref{mfl}). This comparison is shown in Fig. \ref{dens}. 
For both values of $\rho_O$, the ``hybrid'' density profile presents some
ripples near the C$\rightarrow$P cell 
which are damped after around $3\sigma$. Within the P$\rightarrow$C cell the hybrid density profile perfectly 
matches the density within the bulk. 

We also estimated the cost in CPU time of the {\sc usher} algorithm
for the particle insertion (see Ref. \cite{SandraBus} for further details).
As an example, in a fluid with $\rho_c=0.8$, using $\rho_O=0.7$ in Eq. (\ref{mfl}), a
total number of particles of $\sim 10^4$ and an overall simulation CPU
time of $20000$ seconds, the CPU time spent within the particle
insertion/extraction subroutine was about $0.003$ times the CPU time
used by the main force subroutine of the MD code used in this study (which
uses the Verlet list for counting neighbouring particles \cite{allen}).
This excellent performance indicates that the {\sc usher} scheme is
well suited for the insertion protocol in the hybrid scheme when
using Lennard-Jones particles as the fluid of interest.

\section{Shear stress fluctuations and accuracy limits
\label{FL}}

In our scheme, the fluctuating nature of the fluxes introduced into
the C region from P$\rightarrow $C imposes a limitation on our ability
to resolve the flow field, as also arises in experiments and full MD
simulations. This limit is determined by the
signal-to-noise ratio becoming smaller than one.  We shall now
consider the effect of fluctuations of the transversal momentum flux
and provide an estimation of its signal-to-noise ratio in order to
derive the resolution limit of the hybrid scheme. This task requires
consideration of the autocorrelation of the transversal momentum flux or, in other words,
of the $xy$ component of the microscopic stress tensor defined in Eq. (\ref{jp})
\begin{equation}
G(t)=\frac{V_{PC}}{T} \langle j^{\prime}_{p,xy}(t) j^{\prime}_{p,xy}(0)\rangle_{st},
\label{eta}
\end{equation}
where $\langle.\rangle_{st}$ denotes the statistical ensemble average
(assuming ergodicity, we measured it by performing the time average 
over a long enough simulation time) and $j^{\prime}_{p,xy}=j_{p,xy}-\langle j_{p,xy} \rangle_{st}$.
Equation (\ref{eta}) gives a direct relationship between the
amplitude of the stress tensor fluctuations and the shear modulus
$G(0)$
\begin{equation}
\label{eta2}
\langle (j^{\prime}_{p,xy})^2 \rangle_{st}= \frac{T}{V_{PC}} G(0).
\end{equation}
In the case of a LJ fluid the shear modulus is known and is
given by $G(0)=3P-\frac{24}{5} \rho U -2 \rho T$ \cite{Zwa65,Hes97}.
For other interatomic potentials the route described above would
require the evaluation of $G(0)$ from MD simulations since the algebraic relationship
of $G(0)$ with thermodynamic variables is not normally available. 

To evaluate the size of the stress tensor fluctuations we
shall use an equivalent route involving the transport coefficient (shear
viscosity) and the corresponding Green-Kubo relation
\begin{equation}
\eta =\int _{0}^{\infty } G(t) dt.
\label{eta3}
\end{equation}
Let us approximate $G(t)\approx \eta \exp(-t/\tau_G)/\tau_G$, where
$\tau_G$ is a decorrelation time which can be calculated from the
outcome of a simulation by 
$\tau_G = \int_0^{\infty} G(t)dt/G(0)$. Using  Eqs. (\ref{eta})-(\ref{eta2}) one obtains
\begin{equation}
\label{pi1}
\langle (j^{\prime}_{p,xy})^{2}\rangle_{st} =\frac{T\eta }{V_{PC}\tau_G}.
\end{equation}
We note that $\lim_{\tau_G\rightarrow 0} [\exp(-t/\tau_G)/\tau_G] =
2\delta (t)$, where $\delta (t)$ is the Dirac delta function. Hence
the Landau expression for the fluctuation of the stress tensor
$\bm{\Pi}$ is coherently recovered from (\ref{pi1}) in the limit of
the delta correlated process, $\tau_G \rightarrow 0$
\begin{equation}
\label{land}
\langle (\Pi^{\prime}_{xy})^2\rangle_{st}=\frac{2T\eta }{V_{PC}}\delta (t).
\end{equation}
The Landau expression \cite{LandauFL} comes from
a Langevin dynamics approach and is valid for a coarse-grained process in
which the time step (or time between measurements) is much larger than the
decorrelation time $\tau_G$.

Table 2 shows the amplitude of the fluctuations of the microscopic
shear stress $\left<(j^{\prime}_{p,xy})^2\right>$ for several calculations of a
Couette flow at very small shear rate $\gamma \simeq 0.002$. We
confirmed that for these values of $\gamma$ the variance of the shear
stress and velocity are similar to their values at equilibrium.
Results were obtained for both the LJ and the WCA fluid. The numerical
results are in good agreement with the theoretical expression given in
Eq. (\ref{pi1}), where the decay time was measured from simulations, $\tau_G \simeq 0.06$. In
the case of the LJ fluid (for which the value of $G(0)$ is
available) the values of $\left< (j^{\prime}_{p,xy})^2\right>$ obtained from
Eq. (\ref{eta2}) are compared with Eq. (\ref{pi1}) in Table 2.

The standard deviation of the microscopic momentum flux 
averaged over a time interval $\Delta t_{av}=n_{s}\delta t_{s}$ is
given by $S^2=\left<(j^{\prime}_{p,xy})^2\right>/n_s$. We note that for this
relation to hold the time interval between samples $\delta t_{s}$
needs to be somewhat larger than $\tau_G\simeq 0.06$, in order to
ensure a number $n_s$ of statistically independent evaluations of
$j_{p,xy}$.

\begin{table}
\begin{tabular}{|l|l|l|l|c|c|}
\hline 
& & & & Numerical & Theoretical \\
Fluid & $V_{PC}$ &  $T$& $\eta$ & $\left< (j^{\prime}_{p,xy})^2\right>$ &
$\left<(j^{\prime}_{p,xy})^2\right>$ \\
\hline
WCA&  81 &   1 & 1.75 & 0.51 & 0.60 \\
WCA & 173 &  1 & 1.75 & 0.40 & 0.41 \\
WCA & 138 &  1 & 1.75 & 0.33 & 0.38\\
WCA & 338 &  1 & 1.75 & 0.25 & 0.24\\
WCA& 2778 &  1 & 1.75 & 0.08 & 0.06\\
WCA& 2778 &  1 & 1.75 & 0.04 & 0.03\\
LJ& 121.5 & 4.0& 2.12 & 1.09 &1.08 (1.19) \\
LJ& 121.5 & 2.0& 1.90 & 0.66 &0.72 (0.71) \\
LJ& 121.5 & 1.0& 1.75 & 0.39 &0.49 (0.43) \\
\hline
\end{tabular}
\caption{The variance of the microscopic momentum flux 
for $\rho =0.8$. The theoretical results come from Eq. (\ref{pi1})
using $\tau_G=0.06$; those in parentheses are derived from Eq. (\ref{eta2}).}
\end{table}

For the shear flows considered here 
the mean momentum flux can be expressed as 
$\eta \gamma$ and the signal-to-noise ratio is $E\equiv \eta \gamma/S$.
We now demand $E>O(1)$ to obtain conditions on
$V_{PC}$ and $\Delta t_{av}$; using Eq. (\ref{pi1}) one obtains
\begin{equation} 
\label{stn}
  E^{2}= \frac{\gamma^2 \eta}{T} \frac{\tau_G}{\delta t_s} \Delta t_{av} V_{PC}>O(1)
\end{equation}
and using $\delta t_s \geq \tau_G$ one gets,
\begin{equation}
\label{s2}
V_{PC}\Delta t_{av} > \frac{T}{\gamma ^{2}\eta}.
\end{equation}
Equations (\ref{stn}) and (\ref{s2}) give some control over the
amplitude of the fluctuations of the coarse-grained (space-time
averaged) stress tensor.  For instance, in the case of steady state
flows the signal-to-noise ratio can be increased by enlarging
$V_{PC}$, $\Delta t_{av}$, or both.  But if the flow is not
homogeneous in space or if it is time-dependent, $V_{PC}$ and
$\Delta t_{av}$ are bounded above by the requirement of spatial and
temporal flow resolution.  In other words,
to resolve a flow whose shortest characteristic time is $\omega_{\max}^{-1}$
and whose maximum wavenumber is $k_{\max }$, the temporal and spatial
resolution conditions are $\Delta t_{av} \omega_{\max}^{-1} \leq O(0.1)$
and $\Delta x_{PC} k_{\max} \leq O(0.1)$, where $V_{PC}=A_{PC}\Delta
x_{PC}$ (see Fig. \ref{1}). Inserting these conditions into Eq. (\ref{s2}) one gets that
the range of frequencies and wavenumber that can be resolved is 
bounded above by  $k\omega<\gamma^2\eta T^{-1} A_{PC}^{-2}$.

\section{Application to oscillatory shear flow
\label{sec.osc}}
In order to test the applicability of the full hybrid scheme under
unsteady flows, we consider the flow of an incompressible and
isothermal fluid between two parallel walls in relative oscillatory
motion.  In our test flow, the simulation domain is
$0\leq x\leq L_x$ and it is periodic along $y$ and $z$ directions.
The particle domain occupies the region $x<l_P$, and it includes the
LJ liquid and the atomistic wall composed of two layers LJ particles
at $x\leq 0$. The continuum domain comprises the region
$x\in[l_C,L_x]$. The sizes of the simulation domains are within the
nanoscale $L_x \sim 50\sigma$, and $l_P \sim 15\sigma$, while the
width of the handshaking region are $l_P-l_C \sim 5\sigma$.  The flow
is uniquely driven by the oscillatory motion of the $x=L_x$ wall along
the $y$ direction, meaning that the mean pressure is constant
throughout the domain and there are no transfers of mean energy or
mass in the $x$ direction (perpendicular to the P$\rightarrow$C
surface).  The mean flow carries transversal
momentum by diffusion only, and the equation of motion for the
$y$-velocity is $\partial u/\partial t=\nu \partial^2 u/\partial x^2$,
with boundary conditions $u(0,t)=0$ and
$u(L,t)=u_{wall}(t)=u_{\max}\,\sin(\omega t)$.  This equation can be
solved analytically \cite{Sch,Car59,Wij02}.  The flow profile has a
maximum amplitude at the moving wall and the momentum introduced by
its motion penetrates into a fluid layer of width $\delta \sim
\sqrt{\pi\nu/f}$.  Beyond this layer (usually called the viscous layer) the
flow amplitude tends to zero diffusively as it approaches the other
wall held at rest. Therefore, the maximum shear rate attained inside
or near the viscous layer is of order $\gamma \sim u_{\max}/\delta$.
Inserting this relation into the signal-to-noise condition (\ref{s2}),
we find
\begin{equation}
\rho u_{\max}^2 \Delta t_{av} > \pi f^{-1} \left(\frac{k_B T}{V_{PC}}\right),
\label{s3}
\end{equation}
which clearly means that 
in order to attain a signal-to-noise ratio $E$ larger than one, the
mean kinetic energy of the flow integrated over the averaging time
$\Delta t_{av}$ needs to be larger than the net energy due to fluctuations
over a period of the mean flow. We note that at low
enough frequencies, $f>\nu/L^2$, diffusion is the fastest process in
the system (momentum has enough time
to spread mass over the whole domain) and in such cases the correct condition to
guarantee $E<1$ is given by Eq. (\ref{s2}), with $\gamma\sim u_{\max}/L^2$.

In order to resolve the temporal variation of the flow the averaging
time has to be a fraction of the shorter characteristic time of the
flow, which corresponds here to the period of the oscillation,
$f^{-1}$. This condition, $\Delta t_{av} f \geq O(0.1)$, can be
inserted into Eq. (\ref{s3}) to obtain an estimation of the range of
wall velocities that the scheme can resolve: $u_{\max} > 5
\left(\frac{k_B T}{\rho V_{PC}}\right)^{1/2}$.  For $k_B T=1.0$,
$\rho=0.8$ and $V_{PC}=O(100)$ the above inequality yields $u_{\max}>
0.5$. Provided that $\Delta t_{av} f$ is fixed to a
low value, we note that the former condition is independent of
the external frequency  because the shear stress is approximated by $\gamma
\sim u_{\max}/\delta$. This is only valid up to the end of the viscous
layer because $\gamma=\gamma(x)$ becomes very small beyond this layer
($x>\delta$).  Also, using the analytical flow solution \cite{Wij02}
it can be easily shown that for a fixed $u_{\max}$, the local shear
rate within the viscous layer decreases slightly with the frequency. In
fact, the characteristic Reynolds number of the flow goes like
$f^{-1/2}$: $Re^{*}= u_{\max} \delta/\nu \sim u_{\max} f^{-1/2}
\nu^{-1/2}$.  As an aside, the flow considered here (the oscillatory shear
flow in between infinite planes or stream-wise periodic boundary
conditions) is linearly stable \cite{Vog03}.  As shown in a recent
paper, flow instabilities in the form of three-dimensional rolls do
arise at $Re \equiv u_{\max} L_x/ \nu \simeq 500$ only if the zero
mass-flux condition along the stream-wise direction is imposed by 
confinement of lateral walls \cite{Vog03}.  The range of Reynolds number
considered here was such that $Re \leq 200$.

\begin{figure}[h]
\includegraphics{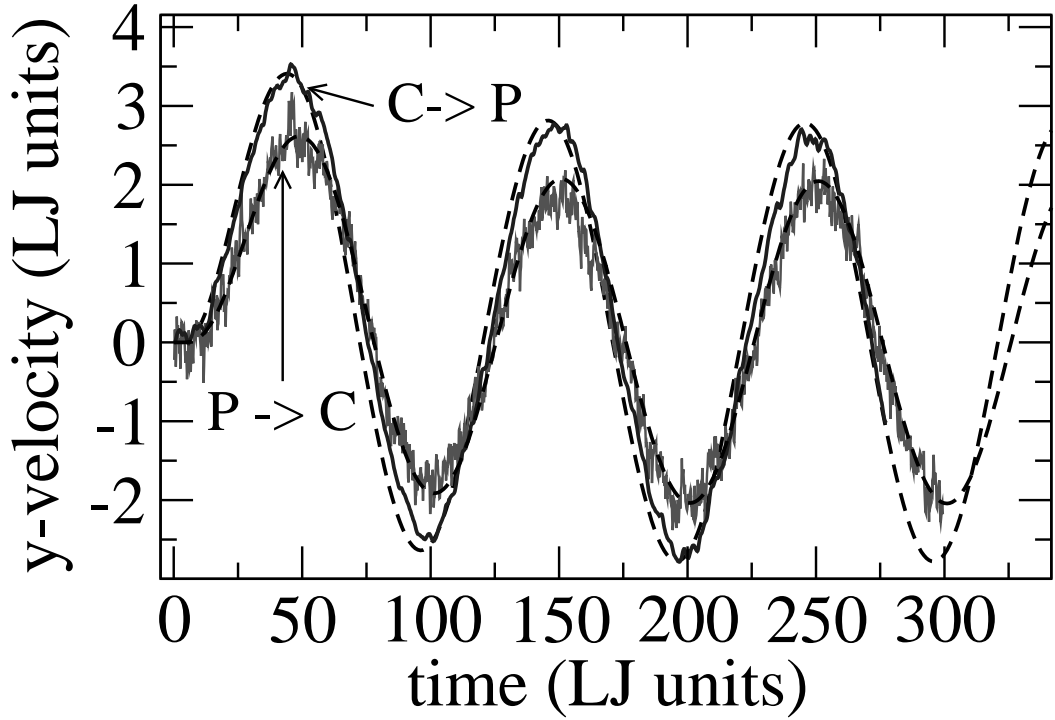}
\caption{ \label{2}
The velocity within the overlapping region for an oscillatory flow with
wall velocity $u_{wall}(t)=u_{\max}\sin(2 \pi f t)$,  
of amplitude $u_{\max}=10$ and frequency $f=0.01$. The
particle domain comprises $x=[0,21]$ and the continuum $x\in [17,30]$;
the extent of the periodic directions is $L_y=L_z=9$. The nanochannel is
filled with a LJ fluid at $\rho =0.8$ and $T=1.0$. 
We plot the instantaneous particle velocity at the P$\rightarrow$C cell
(noisy signal) and the time-averaged velocity over $\Delta t_{av} =1$
at the C$\rightarrow $P cell. Dashed lines correspond to the analytical
solution of the flow.}
\end{figure}

In order to test the hybrid model and the above observations on the
amplitude of fluctuations, we performed oscillatory shear simulations
for values of $u_{max}$ above, near and below  the threshold given by
Eq. (\ref{s3}). In the following calculations the domain geometry is
$L_x=30$ and in the periodic directions $L_y=L_z$ were set to values between
$6$ and $9$.
Calculations for the large amplitude flow are in excellent
agreement with the analytical solution as
illustrated in Fig. \ref{2}, where $u_{\max}=10$, $f=0.01$,
and $\Delta t_{av} =1$. 

\begin{figure}
\includegraphics[width=9cm,totalheight=7cm]{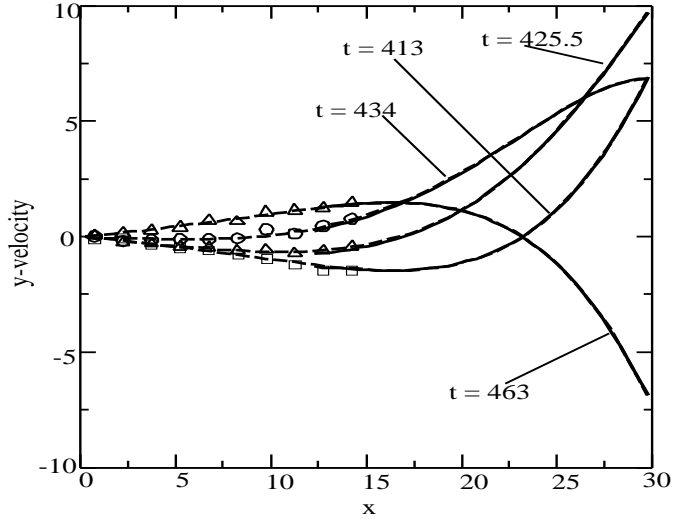}
\caption{ \label{3}Snapshots of the $y$-velocity versus 
the $x$ coordinate at several instants
in time for the same flow as in Fig. \ref{2}. The particle domain comprises
$x\in [0,15]$ and the continuum region $x \in [12, 30]$. The mean
velocities within the P region are calculated within slices of width
$1.6$ and averaged over $\Delta t_{av}=1$; these velocities are plotted 
using different symbols for each time instant. 
We used $\alpha =0.5$ in Eq. (\ref{up}).
The solid lines correspond to the solution within the
continuum region, solved via a finite volume
scheme, and the dashed lines are the analytical solutions. All
quantities are given in reduced LJ units.}
\end{figure}

Figure \ref{3} shows several snapshots of the velocity profiles in the case
of a large amplitude flow ($u_{\max}=10$, $f=0.01$) of a LJ fluid at
density $0.8$ and temperature $T=1.0$. The P region spreads from the
atomic wall at $x=0$ to $x=15$ and the C region comprises
$x=[11.8,30]$. The centres of the P$\rightarrow$C and C$\rightarrow$P
cells were set $3.2$ apart and their width was $1.6$. The outcome of
the hybrid simulation is in very good agreement with the analytical
solution, plotted in dashed lines.

\begin{figure}[h]
\includegraphics[width=9cm,totalheight=9cm]{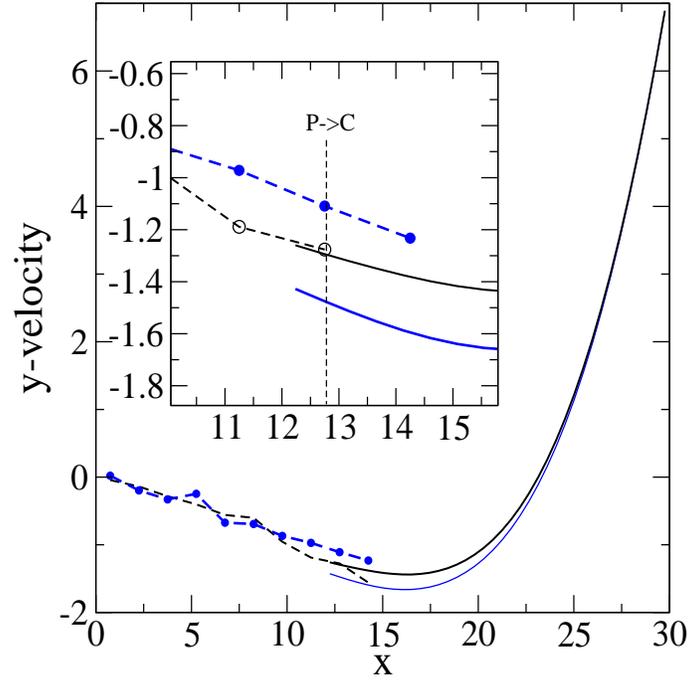}
\caption{ \label{4} Comparison of the velocity profiles obtained with $\alpha=0.5$ 
and $\alpha =0$ in Eq. (\ref{up}). The parameters are
those of Figs. \ref{2} and \ref{3}. Results were obtained with $\Delta t_{av}
=1$. The velocity discontinuity observed in the $\alpha =0$
calculation (about $0.4$) is of the same order of magnitude as the
variance of the mean velocity within the P$\rightarrow $C cell. All
quantities are given in reduced LJ units.}
\end{figure}

The calculations shown in Fig. \ref{3} were performed for a velocity coupling of
$\alpha=0.5$. Figure \ref{4} compares this result with another calculation
using $\alpha=0$. In the latter case, discontinuities of velocity are
clearly observed within the overlapping region, its maximum size, around
$0.35$, being consistent with the size of the velocity (and momentum
flux) fluctuations within the overlapping region.  It is interesting
to note that such discontinuities can be smoothed out by using values
of $\alpha$ as small as $0.2$. In other words, the model is not very
sensitive to the actual value of $\alpha$, provided $\alpha>0$.

\begin{figure}[h]
\includegraphics[width=12cm,totalheight=10cm]{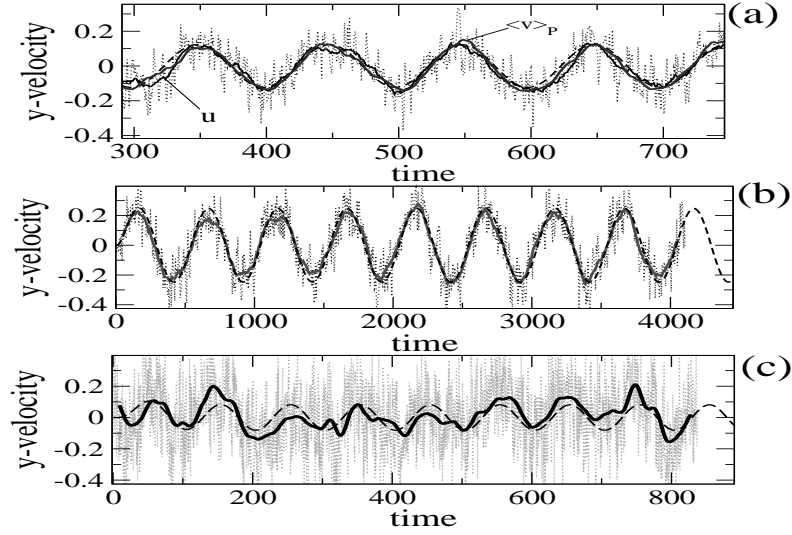}
\caption{\label{5} The mean particle velocity in oscillatory shear flows with
$u_{\max}=0.5$ and $\rho =0.8$. The
molecular dynamics domain comprises $x\in [0,20]$ and the continuum
$x\in [16.36,30]$, the velocity is evaluated within the P$\rightarrow
$C cell (around $x=17.27$).  (a) corresponds to $T=1$ and
$f=0.01$; (b) to $T=1$ and $f=0.002$ and (c) to $T=4$ and
$f=0.01$. Solid lines correspond to the time-averaged mean particle
velocity (in time intervals $\Delta t_{av}=10$), dotted lines to the
instantaneous mean velocity and dashed lines to the analytical
solution of the flow. In (a) we show the time-averaged mean particle
velocity $\langle v \rangle_P$ along with the
continuum velocity $u$ at the same location.}
\end{figure}

Some simulations with the same fluid (LJ at $\rho=0.8$ and $T=1$) but
using a smaller flow amplitude, $u_{\max}=0.5$ are shown in Fig. \ref{5}. The
averaging time was chosen to be $\Delta t_{av} =10$.  The case of
Fig. \ref{5}a corresponds to a frequency of $f=0.01$ and it 
lies approximately at the threshold given by Eq.  (\ref{s3}).  The instantaneous
velocity at the P$\rightarrow$C cell indicates that the noise
amplitude is nearly equal to the flow amplitude; as a consequence, the
corresponding time-averaged velocity (shown in Fig. \ref{5}b) wanders
around the analytical curve, clearly showing the traces of
instantaneous fluctuations.  At smaller frequencies the momentum
spreads over the whole domain and the maximum velocity attained in the neighbourhood of 
the overlapping region is larger (recall that the Reynolds
number depends on $f^{-1/2}$). This explains why the simulation with
$f=0.002$ in Fig. \ref{5}c is in better agreement with the analytical trend
than that for $f=0.01$.  Finally, in Fig. \ref{5}c we present a case with
$u_{\max}=0.5$ but at a larger temperature $T=4$.  The flow
amplitude  is then below the threshold given by Eq.  (\ref{s3}) and
forces arising from thermal fluctuations are larger than those arising from
the mean flow.


\section{Concluding remarks
\label{Con}}

We have presented a hybrid finite volume formulation for the 
P$\rightarrow$C coupling in our atomistic-continuum scheme based on
the exchange of fluxes between both domains.  The scheme has been used
to perform hybrid simulations of the oscillatory shear flow driven by
the harmonic motion of one of the walls of a nano channel.  The particle
domain (P) comprises one of the walls and a part of the adjacent fluid
layer, while the remaining fluid within the slot is described by continuum
fluid dynamics (CFD).  The results show that the flux-coupling
scheme is able to solve unsteady flows up to high characteristic
frequencies corresponding to a Stokes number of $St\equiv 2\pi f
L_x^2/\nu \leq O(100)$. 

A central part of this study has been
focused on the limitations of the scheme due to the effect of the
fluctuations arising from the particle domain. We found that in flows
with low shear rates $\gamma\leq O(10^{-2})$ fluctuations of velocity
and momentum flux may affect the continuity at the overlapping region.
For $\gamma \leq O(10^{2})$ these fluctuations induce differences of
the P and C velocities which are of the same size as the maximum
flow velocity.  This affects the flux-scheme because the momentum flux
injected into the P region is measured from the local velocity
gradient in the C domain. We were able to largely avoid this problem by
implementing a finite volume scheme that uses a hybrid definition of
the velocity gradient, constructed from a linear combination of the
C-velocity and the averaged molecular velocity at the flux boundary
condition for the C domain.  The algorithm derived for time
integration of the C velocity has an extra term that produces a
seamless velocity profile by gently driving the C-velocity towards the
averaged P-velocity.  The strength of the coupling term is
proportional to a coupling parameter $\alpha\in(0,1]$ and the scheme
is not very sensitive to the value of $\alpha$, provided that
$\alpha>0$ (we used $\alpha\sim[0.1-0.5]$).  It is important to
mention that the {\it velocity coupling} term only acts on the C domain and
has no direct effect on the particle dynamics. Such a
procedure avoids the use of Maxwell daemons within the overlapping
region, which appear in schemes that rely on variable-coupling
\cite{Liao98,Tho95}. Another attractive feature of the P$\rightarrow$C
coupling presented here is that it guarantees the velocity
continuity while still preserving the balance of fluxes in the
coarse-grained dynamics.
Indeed, the use of hybrid gradients may be extended to any desired variable.
For instance, in the case of diffusion of mass density, Flekk{\o}y {\em et al.}
\cite{Flek01} used a purely hybrid gradient (i.e., with $\alpha=1$ in
the present notation) to couple random walkers with a finite
difference description of Fickian diffusion.

If the signal-to-noise ratio $E$ of the momentum flux  is smaller than one
the flow cannot be described using a deterministic scheme for the
continuum flow. We have analysed this flow field resolution limit
by calculating $E$ using a route based on the Green-Kubo formalism.
Resolution conditions are written in terms of the shear rate in
Eq. (\ref{s2}) and in terms of the oscillatory-wall velocity in Eq. (\ref{s3}).
Simulations performed near and below the derived resolution limit were shown
to be consistent with the theoretical considerations.
If the flow amplitude is smaller than the threshold given by
Eqs. (\ref{s2}) or (\ref{s3}) the effect of fluctuations 
should also be taken into account by describing the C domain through a 
mesoscopic fluid model that solves the stochastic Navier-Stokes
equations \cite{Pep.tc,Flek_fun}.
In the case of Fickian diffusion, Alexander and coworkers used a
hybrid particle-continuum algorithm to study
the fluctuating hydrodynamic limit \cite{Alex02}. They considered a
hybrid coupling scheme with
both deterministic and stochastic continuum solvers and found that, if the correct stochastic 
representation is used within C, the mean and variance of the density are
correctly coupled within the overlapping region. The authors showed that 
a deterministic hydrodynamic solver
guarantees the coupling of the mean density fields, although in this
case the variance falls exponentially in
the direction of the continuum domain. We observed similar results in our
simulations. The generalisation of our hybrid scheme to largely fluctuating flows 
is feasible and the implementation of a stochastic continuum solver
would not substantially alter the scheme presented here.

\acknowledgements
This research is supported by the European Commission through a Marie Curie
Fellowship to RD-B (HPMF-CT-2001-01210) and by the EPSRC RealityGrid project
GR/R 67699. R.D-B also acknowledges support from project BFM2001-0290.



\newpage

\end{document}